\RequirePackage{fix-cm}
\documentclass[smallextended]{svjour3}       
\smartqed  
\usepackage{graphicx}
\usepackage{amsmath}
\usepackage{graphicx,psfrag,epsf}
\usepackage{enumerate}
\usepackage{natbib}
\usepackage{url} 
\usepackage{subfigure}
\usepackage{ctable}
\usepackage{mathtools}
\usepackage{slashbox}
\usepackage{textcomp} 

\newcommand{\real}{I\hspace{-1.2mm}R}

\def\bx{\boldsymbol{x}}

\def\bz{\boldsymbol{z}}

\def\bgamma{\boldsymbol{\gamma}}
\def\btheta{\boldsymbol{\theta}}

\def\bpi{\boldsymbol{\pi}}

\def\bzero{\boldsymbol{0}}

\journalname{arXiv}
\begin{document}

\title{A new look at the inverse Gaussian distribution}


\author{Antonio Punzo}


\institute{A. Punzo \at
              Department of Economics and Business, University of Catania \\
              Tel.: +39-095-7537640\\
              Fax: +39-095-7537610\\
              \email{antonio.punzo@unict.it}           
}

\date{}

\maketitle

\begin{abstract}
The inverse Gaussian (IG) is one of the most famous and considered distributions with positive support.
We propose a convenient mode-based parameterization yielding the reparametrized IG (rIG) distribution; it allows/simplifies the use of the IG distribution in various statistical fields, and we give some examples in nonparametric statistics, robust statistics, and model-based clustering.
In nonparametric statistics, we define a smoother based on rIG kernels.
By construction, the estimator is well-defined and free of boundary bias.
We adopt likelihood cross-validation to select the smoothing parameter.
In robust statistics, we propose the contaminated IG distribution, a heavy-tailed generalization of the rIG distribution to accommodate mild outliers; they can be automatically detected by the model via maximum \textit{a~posteriori} probabilities.
To obtain maximum likelihood estimates of the parameters, we illustrate an expectation-maximization (EM) algorithm.
Finally, for model-based clustering and semiparametric density estimation, we present finite mixtures of rIG distributions. 
We use the EM algorithm to obtain ML estimates of the parameters of the mixture model.
Applications to economic and insurance data are finally illustrated to exemplify and enhance the use of the proposed models.
\end{abstract}

\section{Introduction}
\label{sec:Introduction}

The inverse Gaussian (IG) is a two-parameter family of distributions with probability density function (pdf) tipically expressed as 
\begin{equation}
f\left(x;\mu,\lambda\right)=\sqrt{\frac{\lambda}{2\pi x^{3}}}\exp\left\{-\frac{\lambda\left(x-\mu\right)^{2}}{2\mu^2x}\right\},\quad 0<x<\infty,
\label{eq:IG distribution}
\end{equation}
where $\mu>0$ is the mean and $\lambda>0$ is the shape parameter, inversely related to the distribution variability.
As well-known \citep[see, e.g.,][Chapter~15]{John:Kotz:cont1:1970}, the pdf in \eqref{eq:IG distribution}, which is seen to be a member of the exponential family, is unimodal, with mode located at
\begin{equation}
\mu\left(\displaystyle\sqrt{1+\frac{9 \mu^2}{4 \lambda^2}}-\frac{3 \mu}{2\lambda}\right),
\label{eq:Mode}
\end{equation} 
and positively skewed, with skewness
\begin{equation}
3\sqrt{\frac{\mu}{\lambda}}.
\label{eq:Skewness}
\end{equation}
For the many attractive properties of this distribution, making it one of the most famous and considered distributions with positive support, see \citet{Twee:Stat:1957} and the review paper by \citet{Folk:Chhi:TheI:1978}.
\citet{Sesh:TheI:2012} provides a detailed list of fields where the IG distribution has been applied with success; see also \citet[][Chapter~15]{John:Kotz:cont1:1970} and \citet[][Chapter~2]{Chhi:TheI:1988}.

To further increase the applicability of the IG distribution, in Section~\ref{sec:Reparameterized IG distribution} we propose a convenient parameterization based on the mode $\theta$ and on a parameter $\gamma$ which is closely related to the distribution variability.
We refer to the resulting distribution as reparametrized IG (rIG).
The adopted parameterization simplifies/allows the use of the IG distribution in some statistical fields, and we give some examples in Section~\ref{sec:Applications in statistic}.
In detail, in Section~\ref{sec:Nonparametric density estimation} we propose a kernel smooth estimator specifically conceived for nonparametric density estimation of positive data.
Kernel functions are chosen from the family of rIG distributions (Section~\ref{subsubsec:Reparametrized inverse Gaussian kernel density estimation}); since their support matches the support of data at hand, no weight is allocated to unrealistic negative values so alleviating the boundary bias issue.
We adopt likelihood cross-validation to select the smoothing parameter (Section~\ref{subsubsec:LCV}).
In Section~\ref{sec:Robustness against mild outliers}, we introduce the contaminated IG distribution, a four-parameter heavy-tailed generalization of the rIG distribution to handle the possible presence of mild outliers.
In addition to the parameters of the rIG distribution, the contaminated IG distribution has one parameter controlling the proportion of outliers and one specifying the degree of contamination (Section~\ref{subsubsec:The contaminated inverse Gaussian distribution}).
We describe an expectation-maximization (EM) algorithm to obtain maximum likelihood (ML) estimates of the parameters (Section~\ref{subsubsec:contaminated IG - ML estimation}).
Advantageously with respect to the rIG distribution, mild outliers are automatically down-weighted in the estimation of $\theta$ and $\gamma$, so providing a robust method of parameter estimation and, once the model is fitted, mild outliers can be directly identified via maximum \textit{a~posteriori} probabilities.
In Section~\ref{subsec:Model-based clustering and semiparametric density estimation}, we define finite mixtures of rIG distributions for semiparametric density estimation and clustering of positive data.
The parameterization of the mixture components in terms of the mode is important in this context if one considers that the multimodality is the most striking feature of a mixture density (cf.~Section~\ref{subsubsec:Mixtures of reparametrized inverse Gaussian distributions}).
In Section~\ref{sec:rIG mixture - Maximum likelihood estimation}, we illustrate an EM algorithm to obtain ML estimates of the mixture parameters.
In order to appreciate the usefulness of the proposed models, in Section~\ref{sec:Real data analysis} we present applications to insurance (Section~\ref{subsec:Bodily injury claims}) and economic (Section~\ref{subsec:Income of Italian households in 1986}) data.
At last, in Section~\ref{sec:Conclusions}, we summarize the key aspects of the proposal, along with future possible extensions.

\section{Reparameterized inverse Gaussian distribution}
\label{sec:Reparameterized IG distribution}

In this section we present our parameterization of the IG distribution (Section~\ref{subsec:The model}) and we give some details about the weighted log-likelihood function (Section~\ref{subsec:Pseudo maximum likelihood estimation}) which can be seen as a generalization of the classical log-likelihood function to be used when sample weights are available.

\subsection{The model}
\label{subsec:The model}

\begin{figure}[!ht]
\centering
\subfigure[$\theta=0.4$\label{fig:IGtheta_04}]
{\includegraphics[width=0.324\textwidth]{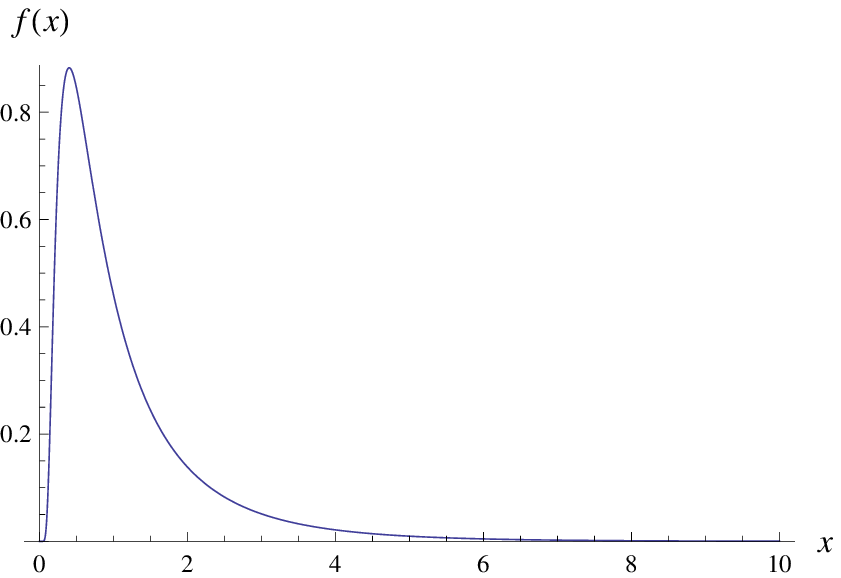}}
\subfigure[$\theta=1$\label{fig:IGtheta_1}]
{\includegraphics[width=0.324\textwidth]{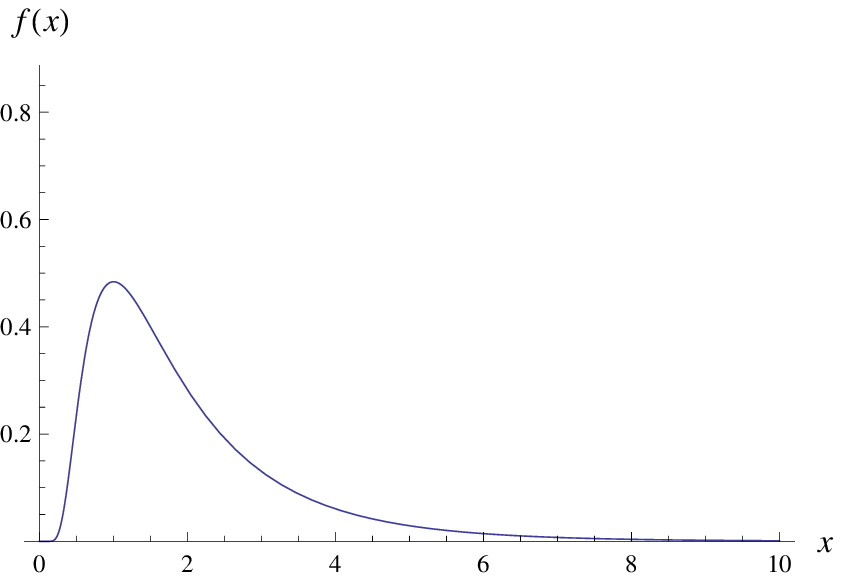}}
\subfigure[$\theta=4$\label{fig:IGtheta_4}]
{\includegraphics[width=0.324\textwidth]{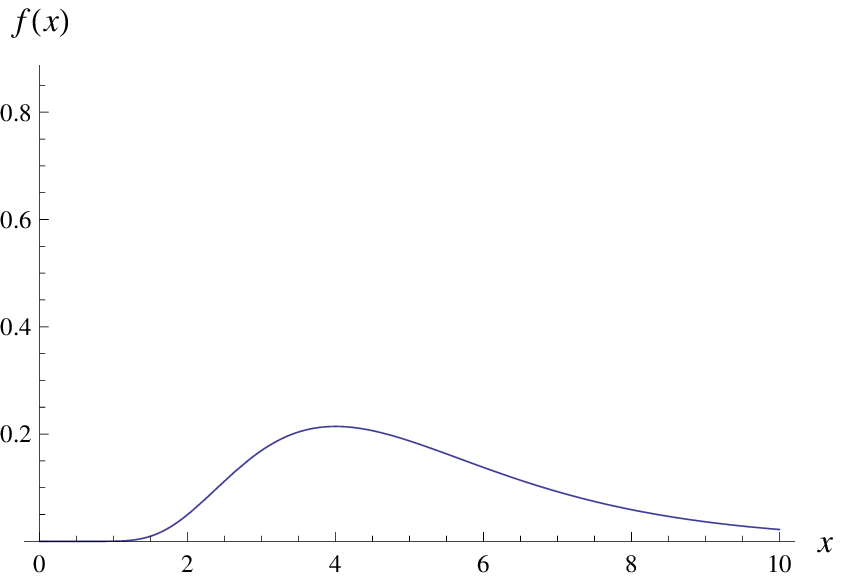}}
\caption{Reparameterized inverse Gaussian pdf in \eqref{eq:MIG distribution} with $\gamma=1$.
\label{fig:IG theta}
}
\end{figure}

\begin{figure}[!ht]
\centering
\subfigure[$\gamma=0.4$\label{fig:IGgamma_04}]
{\includegraphics[width=0.324\textwidth]{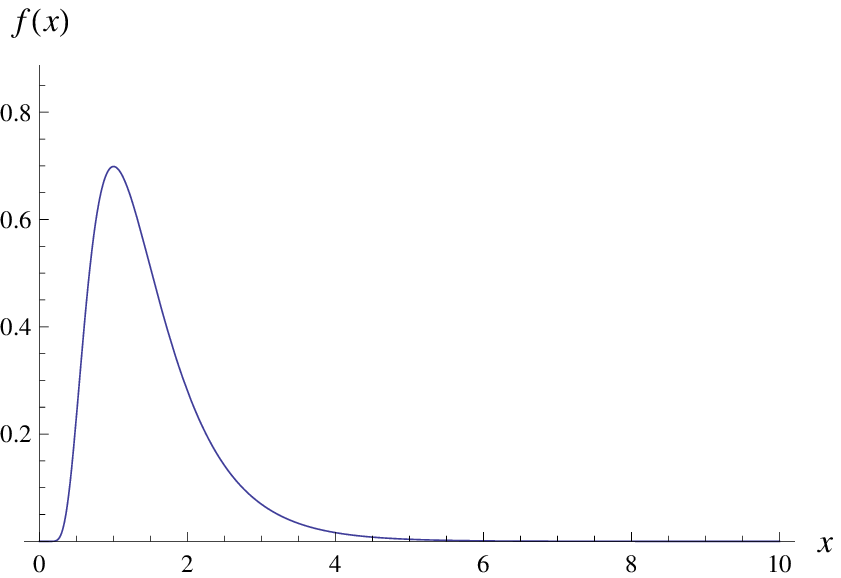}}
\subfigure[$\gamma=1$\label{fig:IGgamma_1}]
{\includegraphics[width=0.324\textwidth]{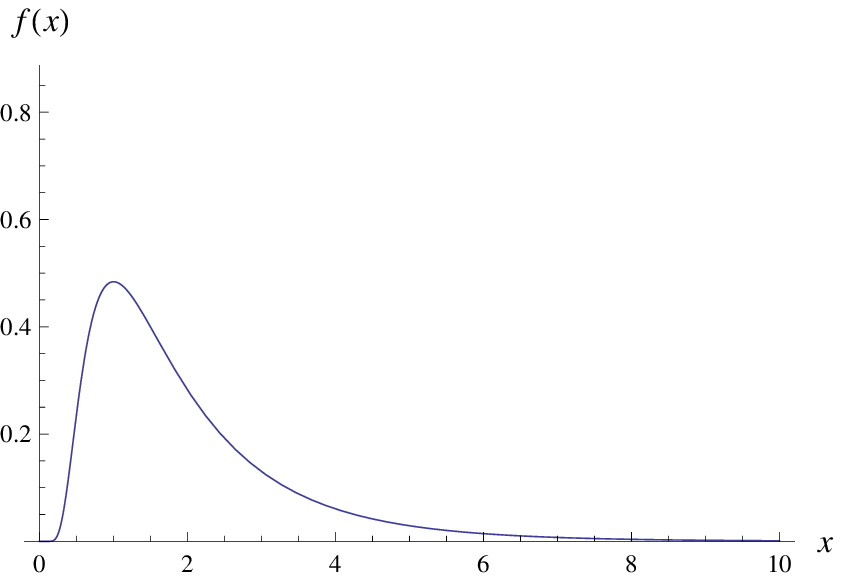}}
\subfigure[$\gamma=4$\label{fig:IGgamma_4}]
{\includegraphics[width=0.324\textwidth]{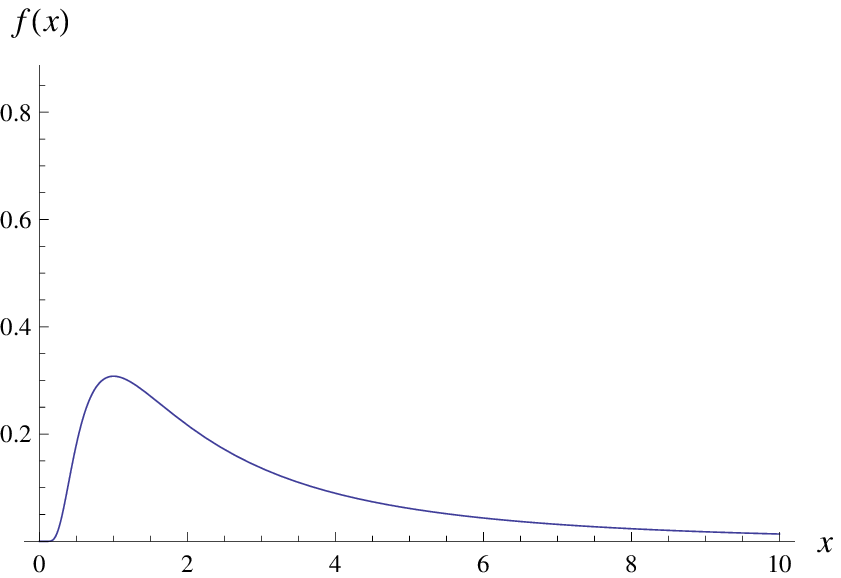}}
\caption{Reparameterized inverse Gaussian pdf in \eqref{eq:MIG distribution} with $\theta=1$.
\label{fig:IG gamma}
}
\end{figure}

The reparametrized IG (rIG) distribution we propose has pdf
\begin{equation}
f\left(x;\theta,\gamma\right)=\sqrt{\frac{\theta\left(3 \gamma + \theta\right)}{2 \pi\gamma x^3}} \exp\left\{-\frac{\left[x-\sqrt{\theta\left(3 \gamma + \theta\right)}\right]^2}{2 \gamma x}\right\},\quad 0<x<\infty, 
\label{eq:MIG distribution}
\end{equation} 
where $\theta,\gamma>0$.
The link between the parameterizations in \eqref{eq:IG distribution} and \eqref{eq:MIG distribution} is 
\begin{equation}
\left\{
\begin{array}{rcl}
\mu &=& \sqrt{\theta\left(3 \gamma + \theta\right)} \\[3mm]
\lambda &=& \displaystyle\frac{\theta\left(3 \gamma + \theta\right)}{\gamma}	
\end{array}
\right.
\qquad\Leftrightarrow\qquad
\left\{
\begin{array}{rcl}
\theta &=& \mu\left(\displaystyle\sqrt{1+\frac{9 \mu^2}{4 \lambda^2}}-\frac{3 \mu}{2\lambda}\right) \\[3mm]	
\gamma &=& \displaystyle\frac{\mu^2}{\lambda}.
\end{array}
\right.
\label{eq:parameterizations link}
\end{equation}
Now, focus on the right-hand side of \eqref{eq:parameterizations link}.
Recalling \eqref{eq:Mode}, the equation on the top guarantees that $\theta$ is the mode of $X$; the effect of varying $\theta$, with $\gamma$ kept fixed, is illustrated in \figurename~\ref{fig:IG theta}. 
The equation on the bottom is chosen so that $\gamma$ is related to the variability of $X$ without making the pdf formulation analytically intractable.
We now try to clarify the role of $\gamma$.
From the standard theory on the IG distribution with pdf given in \eqref{eq:IG distribution}, the variance is $\mu^3/\lambda$ \citep[see, e.g.,][Equation~(15.6)]{John:Kotz:cont1:1970}; thus, thanks to \eqref{eq:parameterizations link}, the variance of the random variable $X$ with pdf \eqref{eq:MIG distribution} is 
\begin{equation*}
\gamma\sqrt{\theta}\sqrt{3\gamma+\theta}.	
\end{equation*}
The last expression, analyzed as a function of $\gamma$, is monotone increasing; consequently, fixed $\theta$, the variability increases in line with the value of $\gamma$, confirming that $\gamma$ governs the spread of $X$.
The effect of varying $\gamma$, the mode $\theta$ kept fixed, is illustrated in \figurename~\ref{fig:IG gamma}.   

\subsection{Maximum weighted likelihood estimation}
\label{subsec:Pseudo maximum likelihood estimation}

Given a sample $x_1,\ldots,x_n$ from the pdf in \eqref{eq:MIG distribution}, the weighted log-likelihood function \citep[see, e.g.,][Chapter 3.4.4]{Skin:Holt:Smit:Anal:1989} related to the rIG distribution is
\begin{equation}
l\left(\theta,\gamma\right) = \sum_{i=1}^n w_i \ln\left[f\left(x_i;\theta,\gamma\right)\right],
\label{eq:weighted loglik}
\end{equation}
where $w_i\geq 0$, $i=1,\ldots,n$, is a given weight.
If $w_1=\cdots=w_n=1$, then the classical log-likelihood function is obtained. 
The use of this function is common when data come from surveys as, for example, in the case of the estimation of the income distribution based on household income data \citep{Graf:Nedy:Muun:Sege:Zins:Para:2011}. 

The first order partial derivatives of \eqref{eq:weighted loglik} with respect to $\left(\theta,\gamma\right)'$ are
\begin{equation}
l'\left(\theta,\gamma\right)=\sum_{i=1}^nw_i\frac{\partial}{\partial \left(\theta,\gamma\right)'} \ln\left[f\left(x_i;\theta,\gamma\right)\right]. 
\label{eq:weighted loglik first derivatives}
\end{equation}
Details about the bidimensional vector of the first order partial derivatives on the right-hand side of \eqref{eq:weighted loglik first derivatives} are given in Appendix~\ref{sec:Partial derivatives of the log pdf of the rIG distribution}. 
Similarly, the second order partial derivatives of $l$ with respect to $\left(\theta,\gamma\right)'$ are
\begin{equation}
l''\left(\theta,\gamma\right)=\sum_{i=1}^nw_i\frac{\partial^2}{\partial \left(\theta,\gamma\right)' \partial \left(\theta,\gamma\right)} \ln\left[f\left(x_i;\theta,\gamma\right)\right].
\label{eq:weighted loglik second derivatives}
\end{equation}
Details about the symmetric $2 \times 2$ matrix of the second order partial derivatives of $\ln\left[f\left(x_i;\theta,\gamma\right)\right]$, on the right-hand side of \eqref{eq:weighted loglik second derivatives}, are given in Appendix~\ref{sec:Partial derivatives of the log pdf of the rIG distribution}.

The values of $\theta$ and $\gamma$ that maximize $l\left(\theta,\gamma\right)$ are the maximum weighted likelihood estimates $\widehat{\theta}$ and $\widehat{\gamma}$ and satisfy the condition
\begin{equation*}
l'\left(\widehat{\theta},\widehat{\gamma}\right)=\bzero.
\end{equation*}
Operationally, we obtain maximization of \eqref{eq:weighted loglik}, with respect to $\theta$ and $\gamma$, by the general-purpose optimizer \texttt{optim()} for \textsf{R} \citep{R:2016}, included in the \textbf{stats} package.
The BFGS algorithm, passed to \texttt{optim()} via the argument \texttt{method}, is used for maximization.

\section{Applications}
\label{sec:Applications in statistic}

In this section we show how our parametrization allows/simplifies the use of the IG distribution in several statistical fields.
We define a smoother based on rIG kernels for nonparametric density estimation (Section~\ref{sec:Nonparametric density estimation}), a contaminated IG distribution for robustness in presence of mild outliers (Section~\ref{sec:Robustness against mild outliers}), and a finite mixture of rIG distributions for clustering/classification and semiparametric density estimation (Section~\ref{subsec:Model-based clustering and semiparametric density estimation}).

\subsection{Nonparametric density estimation} 
\label{sec:Nonparametric density estimation}

Due to their conceptual simplicity and practical and theoretical properties, kernel smoothers are one of the most popular statistical methods for nonparametric density estimation (see, e.g., \citealp{Silv:dens:1986} and \citealp{Wand:Jone:kern:1995}).
Given the random sample $X_1,\ldots,X_n$, these estimators are merely a sum of $n$ (usually symmetric) ``bumps'' (the so-called kernels), with equal weights $1/n$, placed over each observation.
Unfortunately, as stressed in \citet{Chen:beta:1999,Chen:prob:2000}, while using a symmetric kernel is appropriate for fitting distributions with unbounded supports, it is not adequate for distributions with compact or bounded from one end only supports as it causes boundary bias.
The cause of boundary bias is due to the fixed symmetric kernel which allocates weight outside the support when, or especially when (depending from the adopted kernel), smoothing is made near the boundary.

Following the strategy of \citet[][see also \citealp{Punz:Zini:Appr:2012} and  \citealp{Mazz:Punz:Disc:2011,Mazz:Punz:Grad:2012,Mazz:Punz:Usin:2013,Mazz:Punz:DBKG:2014,Mazz:Punz:Biva:2015}]{Punz:disc:2010} in the case of finite discrete supports, in Section~\ref{subsubsec:Reparametrized inverse Gaussian kernel density estimation} we show how a convenient use of rIG kernels automatically permits a solution to boundary bias when the support is $\left(0,\infty\right)$.
Moreover, the resulting estimator is well-defined, that is, the produced estimates satisfy all the fundamental properties of a pdf.
Section~\ref{subsubsec:LCV} suggests an objective selection method to select the smoothing parameter of the proposed density estimator.

\subsubsection{Reparametrized inverse Gaussian kernel density estimation} 
\label{subsubsec:Reparametrized inverse Gaussian kernel density estimation}

Placing a rIG density over each single observation by putting $\theta=X_i$ in \eqref{eq:MIG distribution}, it is possible to consider the following kernel density smoother
\begin{equation}
\widehat{f}\left(x;\gamma\right)=\frac{1}{n}\sum_{i=1}^nf\left(x;\theta=X_i,\gamma\right)=\frac{1}{n}\sum_{i=1}^nk_\gamma\left(x;X_i\right),\quad 0<x<\infty,
\label{eq:kernel estimator}
\end{equation}
where $k_\gamma\left(x;X_i\right)$ and $\gamma$ are the rIG kernel and the smoothing parameter, respectively.
By construction, \eqref{eq:kernel estimator} defines a density function.

Two quantities characterize the nonparametric estimator \eqref{eq:kernel estimator}: the smoothing parameter $\gamma$ and the rIG kernels $k_\gamma\left(x;X_i\right)$.
The former can be considered as smoothing parameter for the following considerations: according to the results of Section~\ref{sec:Reparameterized IG distribution}, if $\gamma$ is chosen too large, then all details, such as modes, may be obscured by $\widehat{f}\left(x;\gamma\right)$.
\textit{Vice versa}, as $\gamma$ becomes small, spurious fine structure becomes visible.
The limit as $\gamma\rightarrow 0^+$ is a sum of $n$ Dirac delta functions (spikes) over the observations; consequently, $\widehat{f}\left(x;\gamma\right)$ converges to the empirical frequency distribution.
As regards the rIG kernels, they obey the fundamental graphical properties of a kernel function.
In detail, they are non-negative, integrate to one, assume their maximum value when $x=X_i$, and are smoothly non-increasing as the point $x$ departs from $X_i$.
The only unconventional property is their skewness: indeed, fixed $\gamma$, the kernel shape changes naturally according to the position where the observation $X_i$ falls (see \figurename~\ref{fig:IG theta}).
In particular, thanks to \eqref{eq:parameterizations link} and recalling \eqref{eq:Skewness}, 
the skewness of the density \eqref{eq:MIG distribution} is 
\begin{equation}
3 \sqrt{\frac{\gamma}{\sqrt{\theta \left(3 \gamma + \theta\right)}}};
\label{eq:reparameterized skewness}
\end{equation}
fixing $\gamma$ in \eqref{eq:reparameterized skewness}, the skewness is a decreasing function of $\theta$.
This characteristic, along with the fact that the support $\left(0,\infty\right)$ of a rIG kernel matches the support of the unknown density, constitutes a natural remedy to the problem of boundary bias.

\subsubsection{The choice of the smoothing parameter $\gamma$}
\label{subsubsec:LCV}

The smoothing parameter $\gamma$ must be specified and has a dramatic effect on the resulting estimate.
Choosing $\gamma$ by trial and error is informative, but it is also convenient to have an objective selection method, and the literature about the topic is vast \citep[see, e.g.,][]{Ston:cros:1974}.
Amongst the existing methods, cross-validation \citep[CV;][]{Ston:cros:1974} is without doubt the most commonly used and the simplest to understand.
Two common CV alternatives are the least squares CV (LSCV; 
\citealp[][pp.~48--49]{Silv:dens:1986}) and the likelihood CV (LCV; 
\citealp[][pp.~52--55]{Silv:dens:1986}).
However, as demonstrated by \citet{Horn:Gart:Like:2006}, LCV generally performs better than LSCV, producing estimates with better fit and less variability, and it is especially beneficial with small sample sizes $n$.
Moreover, LCV has general applicability beyond choosing the smooting parameter in kernel density estimation, having been used for both parameter estimation and model selection \citep[see, e.g.,][]{Ston:cros:1974,Ston:AnAs:1977}. 
The LCV smoothing parameter is chosen by minimizing the score function, suggested by \citet{Duin:OnTh:1976},   
\begin{equation*}
\text{LCV}\left(\gamma\right)=\frac{1}{n}\sum_{i=1}^n \ln \left[\widehat{f}_{-i}\left(x=X_i;\gamma\right)\right]
\end{equation*}
over the possible values of $\gamma$, where $\widehat{f}_{-i}$ is the density estimate in \eqref{eq:kernel estimator} without the data point $X_i$. 
The value of $\gamma$ that minimizes $\text{CV}\left(\gamma\right)$ is referred to as the LCV smoothing parameter, $\widehat{\gamma}_{\text{LCV}}$.
We perform minimization via the \texttt{nlm()} function, of the \textbf{stats} package for \textsf{R}, which carries out a non-linear minimization of $\text{LCV}\left(\gamma\right)$ using a Newton-type algorithm.

\subsection{Robustness against mild outliers} 
\label{sec:Robustness against mild outliers}

Although the IG is one of the most considered distributions with support $\left(0,\infty\right)$, real data are often 
``contaminated'' by outliers --- at one or both ends of the support --- that can affect the estimation of the parameters.
Thus, the detection of outliers, and the development of robust methods of parameter estimation insensitive to their presence, is an important problem.

Outliers can be roughly distinguished into two types \citep[cf.][pp.~79--80]{Ritt:Robu:2015}: mild (also referred to as bad points herein, in analogy with \citealp{Aitk:Wils:Mixt:1980}) and gross.
Mild outliers, on which we focus on, are observations sampled from some population different or even far from the assumed model. 
Such outliers document mainly the difficulty of the specification problem. 
In their presence the statistician is recommended to choose a model flexible enough to accommodate all data points, including the outliers.
The classical choice is to consider heavy-tailed distributions; endowed with heavy tails, they offer the flexibility needed for achieving mild outliers robustness.
Heavy tails are typically obtained by embedding the reference distribution (the IG in our case) in a larger model with one or more additional parameters denoting deviation from the reference distribution due to mild outliers; for a discussion about the concept of reference distribution, see \citet{Davi:Gath:Thei:1993} and \citet{Henn:Fixe:2002}.

By choosing the rIG as reference distribution, in Section~\ref{subsubsec:The contaminated inverse Gaussian distribution} we propose a simple four-parameter contaminated model in order to accommodate all the available data points.
The proposed model is a two-component mixture in which one of the components, with a large prior probability, represents the good points (reference distribution), and the other, with a small prior probability, the same mode, and an inflated parameter $\gamma$, represents the bad points. 
This is a simple theoretical model for the occurrence of bad points and the two additional parameters, with respect to the parameters of the reference rIG distribution, have a direct interpretation in terms of proportion of good points and degree of contamination (a sort of measure of how different bad points are from the bulk of the good points).
Advantageously, the proposed model also allows for automatic detection of bad points via a simple and natural procedure based on maximum \textit{a~posteriori} probabilities.
Note that, as we will detail in Section~\ref{subsubsec:The contaminated inverse Gaussian distribution}, the parameterization of the IG distribution given in \eqref{eq:MIG distribution} is fundamental for the definition of the contaminated model.
We discuss maximum likelihood (ML) estimation of the parameters for the contaminated IG distribution in Section~\ref{subsubsec:contaminated IG - ML estimation} via the adoption of the expectation-maximization (EM) algorithm.

\subsubsection{The contaminated inverse Gaussian distribution}
\label{subsubsec:The contaminated inverse Gaussian distribution}

The pdf of the contaminated IG model is given by
\begin{equation}
p\left(x;\theta,\gamma,\alpha,\eta\right)=\alpha f\left(x;\theta,\gamma\right)+\left(1-\alpha\right) f\left(x;\theta,\eta\gamma\right),\quad 0<x<\infty.
\label{eq:contaminated model}
\end{equation}  
In \eqref{eq:contaminated model}:
\begin{itemize}
	\item $f\left(x;\theta,\gamma\right)$ is the pdf of the rIG, given in \eqref{eq:MIG distribution}, chosen as reference distribution.
	\item $\alpha\in\left(0.5,1\right)$ can be seen as the proportion of good points.
	Note that $\alpha$ is constrained to be greater than 0.5 because, in robust statistics, it is usually assumed that at least half of the observations are good \citep[cf.][p.~250]{Henn:Fixe:2002}.
 	\item $\eta>1$ denotes the degree of contamination and, because of the assumption $\eta>1$, it can be interpreted as the increase in variability due to the bad points with respect to the reference distribution $f\left(x;\theta,\gamma\right)$; hence, it is an inflation parameter.
\end{itemize}
Of course, because the reference distribution $f\left(x;\theta,\gamma\right)$ and the inflated distribution $f\left(x;\theta,\eta\gamma\right)$ have their maximum in $\theta$, this also guarantees that $p\left(x;\theta,\gamma,\alpha,\eta\right)$ will produce a unimodal density with mode $\theta$.
As a limiting case of \eqref{eq:contaminated model}, when $\alpha\rightarrow 1^-$ and $\eta\rightarrow 1^+$, the reference distribution $f\left(x;\theta,\gamma\right)$ is obtained.

An advantage of model~\eqref{eq:contaminated model} is that, once $\theta$, $\gamma$, $\alpha$, and $\eta$ are estimated, say $\widehat{\theta}$, $\widehat{\gamma}$, $\widehat{\alpha}$, and $\widehat{\eta}$, we can establish whether a generic data point, say $x^*$, is either good or bad via the \textit{a~posteriori} probability
\begin{equation}
P\left(\text{$x^*$ is good}\left|\widehat{\theta},\widehat{\gamma},\widehat{\alpha},\widehat{\eta}\right.\right)=\frac{\widehat{\alpha}f\left(x^*;\widehat{\theta},\widehat{\gamma}\right)}{p\left(x^*;\widehat{\theta},\widehat{\gamma},\widehat{\alpha},\widehat{\eta}\right)}.
\label{eq:probability good}
\end{equation}
Based on \eqref{eq:probability good}, $x^*$ will be considered good if $P\left(\text{$x^*$ is good}\left|\widehat{\theta},\widehat{\gamma},\widehat{\alpha},\widehat{\eta}\right.\right)>1/2$, while it will be considered bad otherwise.
The resulting information can be used to eliminate the bad points, if such an outcome is desired \citep{Berk:Bent:Esti:1988}.

\subsubsection{Maximum likelihood estimation: An EM algorithm}
\label{subsubsec:contaminated IG - ML estimation}

In analogy with Section~\ref{subsec:Pseudo maximum likelihood estimation}, estimates of the parameters $\theta$, $\gamma$, $\alpha$, and $\eta$ can be determined by the maximization of the weighted log-likelihood function if sample weights $w_1,\ldots,w_n$ are available in addition to the sample $x_1,\ldots,x_n$ from model~\eqref{eq:contaminated model}.
Details about the four first order partial derivatives of $\ln\left[p\left(x;\theta,\gamma,\alpha,\eta\right)\right]$ are given in Appendix~\ref{sec:Partial derivatives of the log pdf of the contaminated rIG distribution} for the reader interested in this approach. 

Below, to find classical ML estimates of the parameters, we illustrate the use of the EM algorithm \citep{Demp:Lair:Rubi:Maxi:1977}, which is a natural approach for ML estimation when data are incomplete. 
In our case, the source of missing data arises from the fact that we do not know whether the generic data point $x_i$, $i=1,\ldots,n$, is good or bad.
To denote this source of missing data, we use the indicator variables $v_1,\ldots,v_i,\ldots,v_n$, where $v_i=1$ if $x_i$ is good and $v_i=0$ otherwise, $i=1,\ldots,n$. 
Therefore, the complete-data are given by $\left(x_1,v_1\right),\ldots,\left(x_i,v_i\right),\ldots,\left(x_n,v_n\right)$ and the complete-data likelihood, on which the algorithm works on, can be written as
\begin{equation*}
L_c\left(\theta,\gamma,\alpha,\eta\right)=\prod_{i=1}^n\left[\alpha f\left(x_i;\theta,\gamma\right)\right]^{v_i}\left[\left(1-\alpha\right) f\left(x_i;\theta,\eta\gamma\right)\right]^{1-v_i}.
\end{equation*}
Simple algebra yields the following complete-data log-likelihood
\begin{equation}
l_c\left(\theta,\gamma,\alpha,\eta\right)=l_{1c}\left(\alpha\right)+l_{2c}\left(\theta,\gamma,\eta\right),
\label{eq:complete-data log-likelihood}
\end{equation}
where 
\begin{equation}
l_{1c}\left(\alpha\right)=\sum_{i=1}^{n}\left[v_i\ln \alpha+\left(1-v_i\right)\ln \left(1-\alpha\right)\right]
\label{eq:l1c}
\end{equation}
and
\begin{equation}
l_{2c}\left(\theta,\gamma,\eta\right)=\sum_{i=1}^{n}\left[v_i\ln f\left(x_i;\theta,\gamma\right)+\left(1-v_i\right)\ln f\left(x_i;\theta,\eta\gamma\right)\right].
\label{eq:l2c}
\end{equation}
The EM algorithm iterates between two steps, one E-step and one M-step, until convergence.
We implement the EM algorithm in \textsf{R}.
 
\paragraph{E-step}

The E-step, on the $\left(r+1\right)$th iteration of the EM algorithm, requires the calculation of $Q\left(\theta,\gamma,\alpha,\eta\right)$, the current conditional expectation of $l_c\left(\theta,\gamma,\alpha,\eta\right)$.
To do this, we need to calculate $E\left(V_i|x_i;\theta^{\left(r\right)},\gamma^{\left(r\right)},\alpha^{\left(r\right)},\eta^{\left(r\right)}\right)$, where $V_i$ is the random variable related to $v_i$, $i=1,\ldots,n$; this expectation is given by
\begin{equation*}
E\left(V_i|x_i;\theta^{\left(r\right)},\gamma^{\left(r\right)},\alpha^{\left(r\right)},\eta^{\left(r\right)}\right)=\frac{\alpha^{\left(r\right)} f\left(x_i;\theta^{\left(r\right)},\gamma^{\left(r\right)}\right)}{p\left(x_i;\theta^{\left(r\right)},\gamma^{\left(r\right)},\alpha^{\left(r\right)},\eta^{\left(r\right)}\right)} \eqqcolon v_i^{\left(r\right)},
\end{equation*}
which is the posterior probability that $x_i$ is a good point; compare with \eqref{eq:probability good}.
Then, by substituting $v_i$ with $v_i^{\left(r\right)}$ in \eqref{eq:complete-data log-likelihood}, and based on \eqref{eq:l1c} and \eqref{eq:l2c}, we obtain $Q\left(\theta,\gamma,\alpha,\eta\right)=Q_1\left(\alpha\right)+Q_2\left(\theta,\gamma,\eta\right)$. 

\paragraph{M-step}

The M-step on the $\left(r+1\right)$th iteration of the EM algorithm requires the calculation of $\theta^{\left(r+1\right)}$, $\gamma^{\left(r+1\right)}$, $\alpha^{\left(r+1\right)}$, and $\eta^{\left(r+1\right)}$ as the values of $\theta$, $\gamma$, $\alpha$, and $\eta$ that maximize $Q\left(\theta,\gamma,\alpha,\eta\right)$.
The update for $\alpha$ is calculated independently by maximizing 
\begin{equation*}
Q_1\left(\alpha\right)=\sum_{i=1}^{n}\left[v_i^{\left(r\right)}\ln \alpha+\left(1-v_i^{\left(r\right)}\right)\ln \left(1-\alpha\right)\right]
\end{equation*}
with respect to $\alpha$, subject to the constraint on this parameter.
Some simple algebra yields
\begin{displaymath}
\alpha^{\left(r+1\right)}=\max\left\{0.5,\frac{1}{n}\sum_{i=1}^nv_i^{\left(r\right)}\right\}.
\end{displaymath}
The updates of $\theta$, $\gamma$, and $\eta$ are obtained by the maximization of the function
\begin{equation}
Q_2\left(\theta,\gamma,\eta\right)=\sum_{i=1}^{n}\left[v_i^{\left(r\right)}\ln f\left(x_i;\theta,\gamma\right)+\left(1-v_i^{\left(r\right)}\right)\ln f\left(x_i;\theta,\eta\gamma\right)\right].
\label{eq:objfun2 M step}
\end{equation}
For \textsf{R} users, the \texttt{optim()} function, in the \textbf{stats} package, can be used to perform a numerical search of the maximum $(\theta^{\left(r+1\right)},\gamma^{\left(r+1\right)},\eta^{\left(r+1\right)})'$ of the function \eqref{eq:objfun2 M step}.


\subsection{Model-based clustering and semiparametric density estimation} 
\label{subsec:Model-based clustering and semiparametric density estimation}

Finite mixtures of distributions are commonly employed in statistical modeling for two different purposes \citep[][pp.~2--3]{Titt:Smit:Mako:stat:1985}.
In \textit{indirect applications}, they are used as semiparametric competitors of nonparametric density estimation techniques (\citealp[][pp.~28--29]{Titt:Smit:Mako:stat:1985}, \citealp[][p.~8]{McLa:Peel:fini:2000}, and \citealp{Esco:West:Baye:1995}).
On the other hand, in \textit{direct applications}, finite mixture models are considered a powerful device for clustering/classification by assuming that each mixture component represents a group (or cluster) in the original data \citep[see][]{McLa:Basf:mixt:1988}.
A wide range of disciplines can benefit from the application of mixture models, from biology and medicine \citep{Schla:Medi:2009} to economics and marketing \citep{Wede:Kama:Mark:2000}; overviews are given in \citet{McLa:Peel:fini:2000}, \citet{Fruh:Fine:2006}, and \citet{mcnicholas16}.

  
Most of the work published is concerned with mixtures of Gaussian distributions; they are able to approximate arbitrarily well any continuous distribution \citep[see, e.g.,][p.~1]{McLa:Peel:fini:2000}.
Although using Gaussian components in the mixture is in principle appropriate when the theoretical support is $\real$, it is not adequate if the support is $\left(0,\infty\right)$ due to the boundary bias issue discussed in Section~\ref{sec:Nonparametric density estimation}.
A simple remedy is to use mixture components defined on $\left(0,\infty\right)$.
Motivated by this consideration, we suggest using rIG components.
The choice of rIG components is justified, but above all natural, if one thinks that the most striking feature of a mixture density is often that of multimodality.
Indeed, as highlighted in \citet{Titt:Smit:Mako:stat:1985} and \citet{McLa:Basf:mixt:1988}, many papers in applied fields talk not in terms of mixtures but of multimodal distributions; examples are the articles by \citet{Murp:onec:1964} and \cite{BSCS:bimo:1983} referring to bimodality rather than to mixtures.

\subsubsection{Mixtures of reparametrized inverse Gaussian distributions}
\label{subsubsec:Mixtures of reparametrized inverse Gaussian distributions}

The finite mixture of rIG densities can be written as
\begin{equation}
p\left(x;\bpi,\btheta,\bgamma\right)=\sum_{j=1}^k\pi_j f\left(x;\theta_j,\gamma_j\right),\quad 0<x<\infty.
\label{eq:rIG mixture}
\end{equation}
In \eqref{eq:rIG mixture}
\begin{itemize}
	\item $f\left(x;\theta_j,\gamma_j\right)$ is the rIG component density with parameters $\theta_j$ and $\gamma_j$;
	\item $\bpi=\left(\pi_1,\ldots,\pi_k\right)'$ is the vector of mixture weights, with $\pi_j\in\left(0,1\right)$ and $\sum_{j=1}^k\pi_j=1$;
	\item $\btheta=\left(\theta_1,\ldots,\theta_k\right)'$ is the vector of component modes $\theta_j$;
	\item $\bgamma=\left(\gamma_1,\ldots,\gamma_k\right)'$ is the vector of component parameters $\gamma_j$.
\end{itemize}
Thus, there are $3k-1$ unknown parameters to be estimated.
Of course, as also underlined by \citet[][p.~103]{Izen:Mode:2008} and \citet{Bagn:Punz:Fine:2013}, there is no guarantee that $p$ will produce a multimodal density with the same number of modes as there are densities in the mixture; similarly, there is no guarantee that those individual modes $\theta_j$ will remain at the same locations in \eqref{eq:rIG mixture}.
Indeed, the shape of the mixture distribution depends upon both the spacings of the modes and the relative shapes of the component distributions.
Nevertheless, we retain that for well-separated components, the values of $\theta_j$ should accurately approximate the location of the mixture modes.

In terms of indirect applications, model~\eqref{eq:rIG mixture} provides a semiparametric compromise between the single (parametric) rIG density given in \eqref{eq:MIG distribution}, in the case $k=1$, and the nonparametric method of density estimation based on rIG kernels given in \eqref{eq:kernel estimator}, in the case $k=n$. 

\subsubsection{Maximum likelihood estimation: The EM algorithm}
\label{sec:rIG mixture - Maximum likelihood estimation}

As for the contaminated IG distribution, to find ML estimates of the parameters for model~\eqref{eq:rIG mixture} we use the EM algorithm. 
In this case the source of incompleteness, the classical one in the use of mixture models, arises from the fact that for each observation we do not know its component membership; this source, which is especially related to a direct application of the model, is governed by an indicator vector $\bz_i=\left(z_{i1},\ldots,z_{ik}\right)$, where $z_{ij}=1$ if $x_i$ comes from component $j$ and $z_{ij}=0$ otherwise.
The complete-data likelihood can be written as
\begin{equation*}
L_c\left(\bpi,\btheta,\bgamma\right) = \prod_{i=1}^n\prod_{j=1}^k\left[\pi_j f\left(x_i;\theta_j,\gamma_j\right)
\right]^{z_{ij}}.
\end{equation*}
Therefore, the complete-data log-likelihood becomes
\begin{equation}
l_c\left(\bpi,\btheta,\bgamma\right)
=l_{1c}\left(\bpi\right)
+l_{2c}\left(\btheta,\bgamma\right),
\label{eq:rIG mixture - complete-data log-likelihood}
\end{equation}
where
\begin{align}
l_{1c}\left(\bpi\right)=&\sum_{i=1}^{n}\sum_{j=1}^{k}{z}_{ij}\ln \pi_j,\label{eq:lc pi}\\
l_{2c}\left(\btheta,\bgamma\right)=& \sum_{i=1}^n\sum_{j=1}^kz_{ij}\ln\left[
f\left(x_i;\theta_j,\gamma_j\right)\right]. \label{eq:lc lambda nu eta}
\end{align}
E-step and M-step are described below. 

\paragraph{E-step}

The E-step, on the $\left(r+1\right)$th iteration of the EM algorithm, requires the calculation of $Q\left(\bpi,\btheta,\bgamma\right)$, the current conditional expectation of $l_c\left(\bpi,\btheta,\bgamma\right)$.
To do this, we need to calculate 
\begin{displaymath}
E\left(Z_{ij}|x_i;\bpi^{\left(r\right)},\btheta^{\left(r\right)},\bgamma^{\left(r\right)}\right)=\frac{\pi_j^{\left(r\right)}f\left(x_i;\theta_j^{\left(r\right)},\gamma_j^{\left(r\right)}\right)}{p\left(x_i;\bpi^{\left(r\right)},\btheta^{\left(r\right)},\bgamma^{\left(r\right)}\right)}\eqqcolon z_{ij}^{\left(r\right)}.
\end{displaymath}
Then, by substituting $z_{ij}$ with $z_{ij}^{\left(r\right)}$ in \eqref{eq:rIG mixture - complete-data log-likelihood}, and based on \eqref{eq:lc pi} and \eqref{eq:lc lambda nu eta}, we obtain 
\begin{equation}
Q\left(\bpi,\btheta,\bgamma\right)=
Q_1\left(\bpi\right)
+
Q_2\left(\btheta,\bgamma\right).
\label{eq:Q}
\end{equation}

\paragraph{M-step}

The M-step on the $\left(r+1\right)$th iteration of the EM algorithm requires the calculation of $\bpi^{\left(r+1\right)}$, $\btheta^{\left(r+1\right)}$, and $\bgamma^{\left(r+1\right)}$ as the values of $\bpi$, $\btheta$, and $\bgamma$ that maximize $Q\left(\bpi,\btheta,\bgamma\right)$.
As the two terms on the right-hand side of \eqref{eq:Q} have zero cross-derivatives, they can be maximized separately.
Maximizing $Q_1\left(\bpi\right)$ with respect to $\bpi$, subject to the constraints on these parameters, yields
\begin{equation*}
\pi_j^{\left(r+1\right)} = \frac{1}{n} \sum_{i=1}^nz_{ij}^{\left(r\right)}, \quad j=1,\ldots,k.
\end{equation*}
Maximizing $Q_2\left(\btheta,\bgamma\right)$ with respect to $\btheta$ and $\bgamma$ (subject to the constraints on these parameters), is equivalent to independently maximizing each of the $k$ expressions
\begin{equation*}
	Q_{2j}\left(\theta_j,\gamma_j\right) =  \sum_{i=1}^nz_{ij}^{\left(r\right)}\ln\left[
f\left(x_i;\theta_j,\gamma_j\right)\right], \quad j=1,\ldots,k.
\end{equation*}
$Q_{2j}\left(\theta_j,\gamma_j\right)$ is a weighted log-likelihood, with weights $z_{ij}^{\left(r\right)}$, $i=1,\ldots,n$, whose maximization has been discussed in Section~\ref{subsec:Pseudo maximum likelihood estimation}.

\section{Real data analysis}
\label{sec:Real data analysis}

In this section we will show how the rIG-based models, introduced in Section~\ref{sec:Applications in statistic}, act on real data coming from different disciplines.

\subsection{Bodily injury claims}
\label{subsec:Bodily injury claims}

The first example comes from the insurance world.
As well-known insurance data are often positive, right-skewed, and leptokurtic \citep{Ibragimov2015}.
Several parametric families of distributions have been considered in the literature to accommodate these peculiarities, including the Pareto, Weibull, log-normal, and gamma distributions \citep{klugman2012loss}.
However, when insurance data exhibit unusual shapes, such as multiple modes, these distributions may not be a good candidate, as well-argued in \citet{LeeLin2010} and \citet{Jeon2013}.
In these cases, a more flexible modelling framework, such as a mixture modelling framework, is to be preferred.
The flexibility of finite mixtures in accommodating various shapes of insurance data is now widely recognized (\citealp{ChoyChan2003}, \citealp{Bernardi2012}, \citealp{Choy2016}, and \citealp{Maruotti2016}). 
Among them, mixtures of gamma distributions were successfully considered in \citet{Dey1995}, \citet{Wiper2001}, and \citet{Venturini2008}.
As we will see in the analysis below, mixtures of rIG distributions, introduced in Section~\ref{subsec:Model-based clustering and semiparametric density estimation}, represent a valid alternative.

We use insurance data from \citet{Remp:Derr:Mode:2005}, which are also available in the \textbf{CASdatasets} package \citep{CASdatasets:2016} for \textsf{R}.
The sample represents the bodily injury claims from Massachusetts closed in 2001.
We consider the $n=272$ claims that are coded as ``other providers'', thus ignoring potentially fraudulent claims; all numbers are in thousand dollars as in the original paper.

\begin{figure}[!ht]
\centering
\resizebox{0.75\textwidth}{!}{
\includegraphics{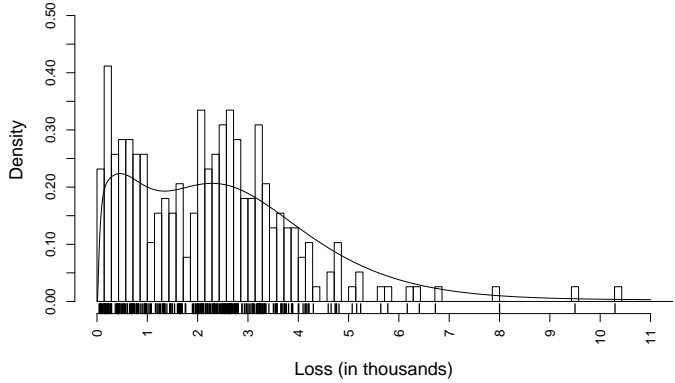} 
}
\caption{
Bodily injury claims.
Histogram together with a rIG-kernel density estimator.
}
\label{fig:histloss}
\end{figure}

The histogram of the data, displayed in \figurename~\ref{fig:histloss}, shows multimodality and right-skewness.
To further explore the characteristics of the empirical pdf, we compute the rIG kernel density estimator introduced in Section~\ref{sec:Nonparametric density estimation}.
The smoothing parameter, selected according to the likelihood cross-validation method discussed in Section~\ref{subsubsec:LCV}, is $\widehat{\gamma}_{\text{LCV}}=0.431$; the corresponding solid curve is superimposed on the histogram in \figurename~\ref{fig:histloss}. 
The nonparametric curve confirms the multimodality suggested by the histogram giving prominence to a clear bimodality.

Motivated by these preliminary findings, we fit mixtures of unimodal gamma distributions \citep{Bagn:Punz:Fine:2013} and mixtures of rIG distributions (introduced in Section~\ref{subsec:Model-based clustering and semiparametric density estimation}) with a number $k$ of mixture components ranging from 1 to 4.
Each model is fitted via the EM algorithm.
To allow for a direct comparison of the competing models, all the algorithms are initialized by providing the initial quantities $\bz_i^{(0)}$, $i=1,\ldots,n$, to the first M-step: 9 times using a random initialization and once with a $k$-means initialization (as implemented by the \texttt{kmeans()} function for \textsf{R}).
The solution maximizing the observed-data log-likelihood among these 10 runs is then selected; see \citet{Dang:Punz:McNi:Ingr:Brow:Mult:2017}.
We select the best value of $k$, as usual in the mixture modelling literature, via the Bayesian information criterion \citep[BIC;][]{Schw:Esti:1978}.
Even though the regularity properties needed for the development of the BIC are not satisfied by mixture models \citep{Keri:Esti:1998,Keri:Cons:2000}, it has been used extensively (see, e.g., \citealp{Dasg:Raft:Dete:1998} and \citealp{Fra:Raf:2002}) and performs well in practice. 
We compute the BIC as
\begin{equation*}
\text{BIC}=2l\left(\widehat{\bpi},\widehat{\btheta},\widehat{\bgamma}\right)-\left(3k-1\right)\ln n,
\end{equation*}
where $l\left(\widehat{\bpi},\widehat{\btheta},\widehat{\bgamma}\right)$ is the maximized (observed-data) log-likelihood. 
Note that, Bayes factors can be used to compare models that are not nested, and the BIC approximation thereto holds when models are not nested (cf.~\citealp{Raft:Baye:1995}).

\begin{figure}[!ht]
\centering
\resizebox{0.75\textwidth}{!}{
\includegraphics{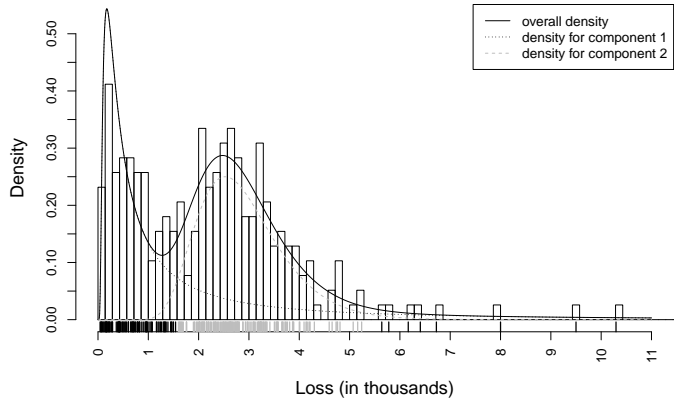} 
}
\caption{
Bodily injury claims.
Histogram together with the fitted mixture of $k=2$ rIG densities.
Dotted lines show the component densities multiplied by the corresponding weights.  
Black and gray are used for observations in group 1 and group 2, respectively, as classified by the fitted model.
}
\label{fig:histmixtIG}
\end{figure}

\tablename~\ref{tab:Students results} shows the obtained BIC values.
\begin{table}[!ht]
\caption{
Bodily injury claims.
BIC values for the fitted models.
Bold numbers refer to the best value of $k$ for each model.  
}
\label{tab:Students results}
\centering
\begin{tabular}{c c rrrr}
  \toprule
  \backslashbox{model}{$k$} && \multicolumn{1}{c}{1} & \multicolumn{1}{c}{2} & \multicolumn{1}{c}{3} & \multicolumn{1}{c}{4} \\ 
  \midrule
mixt. of gamma pdfs && -1093.122 & -1066.879 & \textbf{-1033.998} & -1049.733 \\ 
mixt. of rIG pdfs   && -1169.075 & \textbf{-1026.641} & -1031.266 & -1046.069 \\ 
   \bottomrule
\end{tabular}
\end{table}
The BIC suggests $k=3$ components for mixtures of gamma distributions and $k=2$ components for mixtures of rIG distributions.
These results confirm the observation that a single ($k=1$) parametric model -- gamma or rIG in our case -- is unable to represent the distribution of the bodily injury claims.
Overall, the best model is the mixture of two rIG distributions; its estimated parameters are given in \tablename~\ref{tab:Bic estimated parameters}, while its graphical representation is displayed, via a solid line, in \figurename~\ref{fig:histmixtIG}, with dotted curves showing the component densities multiplied by the corresponding estimated weights $\widehat{\pi}_j$, $j=1,2$.
Group membership of the observations is represented by ticks of different colors (black for group 1 and gray for group 2) on the $x$-axis.
\begin{table}[!ht]
\caption{
Bodily injury claims.
Estimated parameters for the mixture of two rIG distributions.
}
\label{tab:Bic estimated parameters}
\centering
\begin{tabular}{c c rrr}
  \toprule
  \backslashbox{component $j$}{estimates} && \multicolumn{1}{c}{$\widehat{\pi}_j$} & \multicolumn{1}{c}{$\widehat{\theta}_j$} & \multicolumn{1}{c}{$\widehat{\gamma}_j$} \\ 
  \midrule
1 && 0.507 & 0.175 & 11.901 \\ 
2 && 0.493 & 2.527 & 0.262 \\ 
   \bottomrule
\end{tabular}
\end{table}

This application emphasizes the importance of the mode-parameterization, which immediately gives an idea of the location, on the $x$-axis, of the losses with the highest probability (see the third column of \tablename~\ref{tab:Bic estimated parameters}).
In particular, the first mode suggests that a loss of 175 dollars is the most likely for this dataset.
Moreover, the estimated modes can be used to facilitate comparisons across space and time of the two losses more representative of the distribution.

\subsection{Income of Italian households in 1986}
\label{subsec:Income of Italian households in 1986}

The second example comes from the economic literature and it is related to the estimation of the income distribution.
Information from such estimation is used to measure welfare, inequality and poverty, to assess changes in these measures over time, and to compare measures across countries, over time and before and after specific policy changes, designed, for example, to alleviate poverty.
Thus, the estimation of the income distribution is of central importance for assessing many aspects of the well being of society \citep[see][for further considerations]{Silb:Hand:2012}.

The income distribution has been estimated both parametrically and nonparametrically \citep[see, e.g.,][]{Chot:Grif:Esti:2008}.
Parametric estimation is convenient because it facilitates subsequent inferences about inequality and poverty measures based on the estimated income distribution parameters.
A large number of alternative parametric models have been suggested in the literature for estimating the income distribution \citep[see][for a survey]{Klei:Kotz:Stat:2003}.
As well documented in \citet{Dagu:ANew:2008}, a convenient parametric model should be: defined on a strictly positive support, unimodal, and positively skewed; moreover, all the parameters of the specified model should have a well-defined economic meaning and, following a principle of parsimony, the model should make use of the smallest possible number of parameters for adequate and meaningful representation.
Unfortunately, as emphasized by \citet{VanP:Hage:VanE:TheI:1983}, \citet{Fese:Robu:1993}, and \citet{Cowe:Vict:Robu:1996}, real income data are often ``contaminated'' by outliers (bad incomes) that affect the estimation of the parameters for the chosen model.
This in turn will affect the inequality measure computed from the estimated parameters.
As we will see in the analysis below, the contaminated IG distribution can be a remedy to this problem.

\begin{figure}[!ht]
\centering
\resizebox{0.75\textwidth}{!}{
\includegraphics{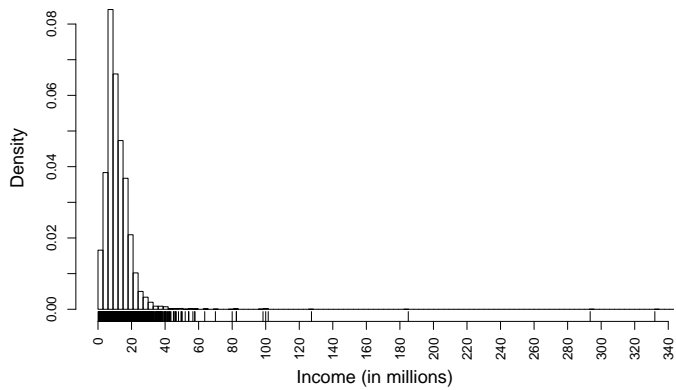} 
}
\caption{
Income of Italian households in 1986.
Weighted histogram.
}
\label{fig:histIncome}
\end{figure}

We use incomes of Italian households, for 1986, obtained from the Luxembourg Income Study (LIS) database (\url{http://www.lisdatacenter.org/}).
The data analyzed here are $n = 6016$ household incomes with corresponding sample weights. 
The weighted histogram of the data, obtained via the function \texttt{wtd.hist()} of the \textbf{weights} package \citep{weights:2016} for \textsf{R}, is displayed in \figurename~\ref{fig:histIncome}.
Although, as expected, the histogram highlights unimodality and positive skewness, some spurious very high incomes appear (see the ticks on the $x$-axis) yielding an heavier right tail.

Motivated by these considerations, we fit the rIG and the contaminated IG distributions to data at hand.
Their nested relationship 
guarantees that the contaminated IG distribution will fit the data at least as well as the rIG distribution. 
However, this superiority could not be statistically significant. 
Thanks to the nested relationship between the competing models, a natural way to compare their goodness-of-fit consists of using the likelihood-ratio (LR) statistic
\begin{displaymath}
	\text{LR} = 2\left[l\left(\widehat{\theta},\widehat{\gamma},\widehat{\alpha},\widehat{\eta}\right)-l\left(\widehat{\theta},\widehat{\gamma}\right)\right],
\end{displaymath}
where $l\left(\widehat{\theta},\widehat{\gamma},\widehat{\alpha},\widehat{\eta}\right)$ and $l\left(\widehat{\theta},\widehat{\gamma}\right)$ are the maximized (observed-data) log-likelihoods for the contaminated and uncontaminated IG models, respectively.
Under the null hypothesis that the true underlying model is the restricted one (the rIG in our case), versus the alternative that the true underlying model is the more complex one (the contaminated IG in our case), 
LR is asymptotically distributed as a $\chi^2$ with two degrees of freedom, corresponding to the difference in the number of free parameters between the null and the alternative model.
Thus, from a practical point of view, the degrees of freedom can be seen as the gain in parsimony that could be obtained using the model under the null instead of the model under the alternative.
With data at hand, the LR statistic assumes value 59.455, and the resulting $p$-value is $1.229\times 10^{-13}$, which leads to the rejection of the null, in favor of the alternative, at any reasonable significance level.

\begin{figure}[!ht]
\centering
\resizebox{0.75\textwidth}{!}{
\includegraphics{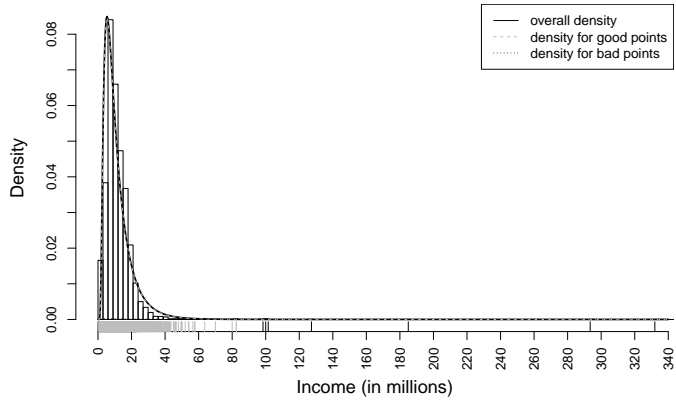} 
}
\caption{
Income of Italian households in 1986.
Weighted histogram together with the fitted contaminated IG density (solid line).
Dotted lines show the densities for good and bad incomes multiplied by the corresponding weights.
Gray and black are used for good and bad incomes, respectively, as classified by the fitted model.
}
\label{fig:histIncomeContaminated}
\end{figure}

The estimated parameters for the contaminated IG distribution are $\widehat{\theta}=5.389$, $\widehat{\gamma}=6.179$, $\widehat{\alpha}=0.991$, and $\widehat{\eta}=15.726$.
The estimated value of $\alpha$ indicates that about the 9\textperthousand\ of the incomes can be considered as bad according to the fitted model, with $\widehat{\eta}$ giving the degree of badness (measure of how far the bad incomes are from the bulk of the data). 
The corresponding estimated curve is represented, via a solid line, in \figurename~\ref{fig:histIncomeContaminated}, along with the weighted histogram; dotted curves show the densities for good and bad incomes multiplied by the corresponding estimated weights $\widehat{\alpha}$ and $1-\widehat{\alpha}$.
Maximum \textit{a~posteriori} classification of incomes, as good or bad, is represented by ticks of different colors (gray for good incomes and black for bad incomes) on the $x$-axis.

\begin{figure}[!ht]
\centering
\resizebox{0.75\textwidth}{!}{
\includegraphics{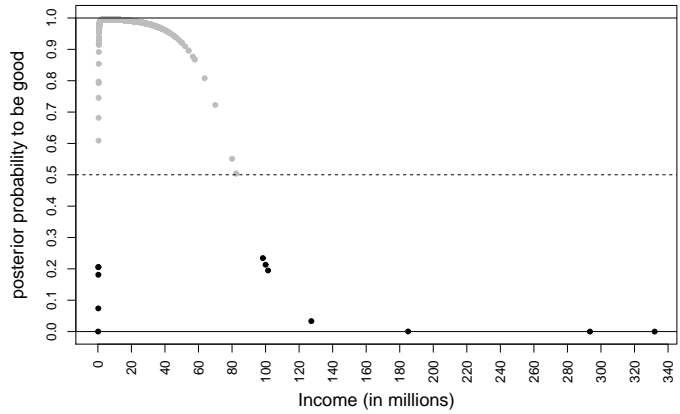} 
}
\caption{
Income of Italian households in 1986.
Estimated posterior probabilities to be good incomes.
Gray and black are used for good and bad incomes, respectively, as classified by the fitted model.
}
\label{fig:postprobIncome}
\end{figure}

\figurename~\ref{fig:postprobIncome} reports, for each income $x_i$, the estimated posterior probability in \eqref{eq:probability good} to be good, $i=1,\ldots,n$; as we can see, the farther the income is from the bulk of the data, as represented by the mode $\widehat{\theta}$, the lower is its probability to be a good income.
Such probability is also related to the down-weighting of bad incomes in the estimation of the model parameters, and this is an important aspect for robust estimation (see \citealp{Punz:McNi:Robu:2016} for a discussion about this topic with reference to the mixture of contaminated normal distributions). 

\section{Conclusions}
\label{sec:Conclusions}

A mode-based parameterization of the inverse Gaussian (IG) distribution was suggested.
It yielded the reparametrized IG (rIG) distribution.
It was used to define three different models to be applied for positive data: a rIG kernel smoother for nonparametric density estimation (Section~\ref{sec:Nonparametric density estimation}), a contaminated IG distribution for robust density estimation (Section~\ref{sec:Robustness against mild outliers}), and a finite mixture of rIG distributions for clustering and semiparametric density estimation (Section~\ref{subsec:Model-based clustering and semiparametric density estimation}).
The real data applications illustrated in Section~\ref{sec:Real data analysis} showed the usefulness of the proposed models.

However, the applicability of our parameterization is not restricted to the models discussed above.
For example, the rIG could be used as distribution of the error term in modal linear regression (\citealp{yao2014new}); the modal linear regression models the conditional mode of a response $Y$ given a set of predictors $\bx$ as a linear function of $\bx$.
Also, in the fashion of \cite{Punz:McNi:Robu:2016}, contaminated IG distributions may be used as components in the definition of a finite mixture model; see also \citet{Punz:Mazz:McNi:ContaminatedMixt:2017}, \citet{Punz:McNi:Robu:2017}, \citet{Punz:Maru:Clus:2016}, and \citet{Maru:Punz:Mode:2016}. 
Finally, in reliability theory, the parameterization with respect to the mode may simplify the formulation of the hazard rate, related to the IG distribution \citep[cf.][Chapter~5.3]{Sesh:TheI:2012}.

\appendix

\footnotesize

\section{Partial derivatives of the log pdf of the rIG distribution}
\label{sec:Partial derivatives of the log pdf of the rIG distribution}

The first order partial derivatives with respect to $\theta$ and $\gamma$, of the logarithm of the pdf in \eqref{eq:MIG distribution}, are
$$
	\frac{\partial \ln\left[f\left(x;\theta,\gamma\right)\right]}{\partial \theta} =
	-\frac{3}{2 x}-\frac{\theta }{x \gamma }+\frac{1}{3 \gamma +\theta }+\frac{3 \gamma }{2 \theta  (3 \gamma +\theta )}+\frac{\sqrt{\theta}}{2 \gamma  \sqrt{3\gamma + \theta}}+\frac{\sqrt{3\gamma + \theta}}{2 \gamma  \sqrt{\theta}}  
$$
and
$$
\frac{\partial \ln\left[f\left(x;\theta,\gamma\right)\right]}{\partial \gamma} =
\frac{x}{2 \gamma^2}+\frac{\theta ^2}{2 x \gamma^2}-\frac{\theta }{2 \gamma  (3 \gamma +\theta )}+\frac{3 \sqrt{\theta}}{2 \gamma  \sqrt{3\gamma + \theta}}-\frac{\sqrt{\theta\left(3\gamma + \theta\right)}}{\gamma^2}. 
$$
The second order partial derivatives are
$$	
\frac{\partial^2 \ln\left[f\left(x;\theta,\gamma\right)\right]}{\partial \theta^2} =
	-\frac{1}{4} \left(\frac{4}{x \gamma } + \frac{2}{\theta ^2} + \frac{2}{(3 \gamma +\theta )^2} + \frac{9 \gamma}{\theta ^{3/2}(3 \gamma +\theta)^{3/2}}\right),  
$$
\begin{align*}
	\frac{\partial^2 \ln\left[f\left(x;\theta,\gamma\right)\right]}{\partial \theta \partial \gamma} 
	&= 
	\frac{\partial^2 \ln\left[f\left(x;\theta,\gamma\right)\right]}{\partial \gamma \partial \theta} 
	\\
	&=  
	\frac{\theta }{x \gamma^2}+\frac{-27 \gamma ^3-30 \gamma  \theta ^2-4 \theta ^3-3 \gamma^2 \left[21 \theta +2 \sqrt{\theta\left(3\gamma + \theta\right)}\right]}{4 \gamma^2 \sqrt{\theta} (3 \gamma +\theta )^{5/2}},
\end{align*}
and
\begin{align*}
	\frac{\partial^2 \ln\left[f\left(x;\theta,\gamma\right)\right]}{\partial \gamma^2} 
	=&
		-\frac{x}{\gamma ^3}-\frac{\theta ^2}{x \gamma ^3}+\frac{3 \theta }{2 \gamma  (3 \gamma +\theta )^2}-\frac{9 \sqrt{\theta}}{4 \gamma  (3 \gamma +\theta )^{3/2}}+\frac{\theta }{2 \gamma^2 (3 \gamma +\theta )}\\ 
		& -\frac{3 \sqrt{\theta}}{\gamma^2 \sqrt{3\gamma + \theta}} + \frac{2 \sqrt{\theta\left(3\gamma + \theta\right)}}{\gamma^3}.
\end{align*}  

\section{First partial derivatives of the log pdf of the contaminated IG distribution}
\label{sec:Partial derivatives of the log pdf of the contaminated rIG distribution}

The first order partial derivatives with respect to $\theta$, $\gamma$, $\alpha$, and $\eta$ of the logarithm of the pdf in \eqref{eq:contaminated model}, are
\begin{align*}
	& \frac{\partial \ln\left[p\left(x;\theta,\gamma,\alpha,\eta\right)\right]}{\partial \theta} = \frac{1}{2 \sqrt{2\pi} \gamma^2 x^4 p\left(x;\theta,\gamma,\alpha,\eta\right)}\times\\
	& \quad 
	\left[\frac{
\alpha  
\left(3 \gamma + 2 \theta\right) 
\left\{
x \left[\gamma + \sqrt{\theta\left(3 \gamma + \theta\right)}\right]
- 
\theta\left(3 \gamma + \theta\right) 
\right\}
\exp\left\{-\frac{\left[x-\sqrt{\theta\left(3 \gamma + \theta\right)}\right]^2}{2 x \gamma}\right\} 
}{
\sqrt{\frac{\theta\left(3 \gamma + \theta\right)}{x^3 \gamma}}
} \right.+ \\
	& \quad
\left.
\frac{
\left(1-\alpha\right) 
\left(3 \eta\gamma + 2 \theta\right) 
\left\{
x \left[\eta\gamma + \sqrt{\theta\left(3\eta\gamma + \theta\right)}\right]
-
\theta\left(3\eta\gamma + \theta\right) 
\right\}
\exp\left\{-\frac{\left[x-\sqrt{\theta\left(3\eta\gamma + \theta\right)}\right]^2}{2 x \eta\gamma}\right\} 
}{\eta^2 \sqrt{\frac{\theta \left(3\eta\gamma + \theta\right)}{x^3 \eta\gamma}}}
\right],
\end{align*}
\begin{align*}
& \frac{\partial \ln\left[p\left(x;\theta,\gamma,\alpha,\eta\right)\right]}{\partial \gamma} = \frac{\theta}{2 \sqrt{2\pi} \gamma^3 x^4 p\left(x;\theta,\gamma,\alpha,\eta\right)}\times \\
	& \quad\qquad
	\left[
\frac{
x^2\left(3 \gamma + \theta\right) 
+ 
\theta^2\left(3 \gamma + \theta\right) 
- 
x \left[
\gamma\theta + 3 \gamma \sqrt{\theta  \left(3 \gamma + \theta\right)} + 2 \theta  \sqrt{\theta  \left(3 \gamma + \theta\right)}
\right]
}{
\sqrt{\frac{\theta\left(3 \gamma + \theta\right)}{x^3 \gamma}}
} \right. \times  
\\
& \quad\qquad
\alpha \exp\left\{-\frac{\left[x-\sqrt{\theta\left(3 \gamma + \theta\right)}\right]^2}{2 x \gamma}\right\} 
+ 
\left(1-\alpha\right) 
\exp\left\{-\frac{\left[x-\sqrt{\theta\left(3\eta\gamma + \theta\right)}\right]^2}{2 x \eta\gamma}\right\}\times \\
	& \quad\qquad
	\left. 
\frac{
x^2\left(3\eta\gamma + \theta\right) + \theta^2 \left(3\eta\gamma + \theta\right) 
- 
x \left[2 \theta  \sqrt{\theta  \left(3\eta\gamma + \theta\right)} + \eta\gamma  \left(\theta + 3 \sqrt{\theta\left(3\eta\gamma + \theta\right)}\right)\right]
}{
\eta^2 \sqrt{\frac{\theta\left(3\eta\gamma + \theta\right)}{x^3 \eta\gamma}}
}
\right],
\end{align*}
\begin{align*}
	& \frac{\partial \ln\left[p\left(x;\theta,\gamma,\alpha,\eta\right)\right]}{\partial \alpha} = \\
	&\qquad \frac{
	\sqrt{\frac{\theta\left(3\gamma + \theta\right)}{x^3 \gamma}}
	\exp\left\{-\frac{\left[x-\sqrt{\theta\left(3\eta\gamma + \theta\right)}\right]^2}{2 x \eta\gamma}\right\} 
	-
	\sqrt{\frac{\theta\left(3\eta\gamma + \theta\right)}{x^3 \eta\gamma }}
	\exp\left\{-\frac{\left[x-\sqrt{\theta\left(3 \gamma + \theta\right)}\right]^2}{2 x \gamma}\right\} 
	}{
	\alpha  
	\sqrt{\frac{\theta\left(3\gamma + \theta\right)}{x^3 \gamma}}
	\exp\left\{-\frac{\left[x-\sqrt{\theta\left(3\eta\gamma + \theta\right)}\right]^2}{2 x \eta\gamma}\right\} 
	+
	\left(1-\alpha\right) 
	\sqrt{\frac{\theta\left(3\eta\gamma + \theta\right)}{x^3 \eta\gamma }}
	\exp\left\{-\frac{\left[x-\sqrt{\theta\left(3 \gamma + \theta\right)}\right]^2}{2 x \gamma}\right\} 
	},
\end{align*}
and
\begin{align*}
	&\frac{\partial \ln\left[p\left(x;\theta,\gamma,\alpha,\eta\right)\right]}{\partial \eta} =
	\frac{\exp\left\{-\frac{\left[x-\sqrt{\theta\left(3 \gamma + \theta\right)}\right]^2}{2 x \gamma}\right\} 
	}{2 x^4 \gamma^2 \eta ^3 
	\frac{\theta  (3 \eta\gamma + \theta )}{\sqrt{x^3 \eta\gamma}}}\times\\ 
	& \qquad\frac{
	\left(1-\alpha\right)
	\left(
	\left(x^2+\theta^2\right)\left[\theta\left(3\eta\gamma + \theta\right)\right]^{3/2}
	- 
	x \theta^2 
	\left\{9 \gamma^2 \eta^2+2 \theta^2+\eta\gamma  \left[9 \theta +\sqrt{\theta\left(3\eta\gamma + \theta\right)}\right]\right\}
	\right)
	}{
	\alpha  
	\sqrt{\frac{\theta\left(3\gamma + \theta\right)}{x^3 \gamma}}
	\exp\left\{-\frac{\left[x-\sqrt{\theta\left(3\eta\gamma + \theta\right)}\right]^2}{2 x \eta\gamma}\right\} 
	+
	\left(1-\alpha\right) 
	\sqrt{\frac{\theta\left(3\eta\gamma + \theta\right)}{x^3 \eta\gamma }}
	\exp\left\{-\frac{\left[x-\sqrt{\theta\left(3 \gamma + \theta\right)}\right]^2}{2 x \gamma}\right\}
	}.
\end{align*}

\bibliographystyle{spbasic}

\end{document}